\documentclass[sigconf,authorversion]{acmart}
\title{Pretraining on Call Graphs: When Binary Analysis Tasks Profit From Context}

\author{Samuel Valenzuela}
\affiliation{
  \institution{Ludwig-Maximilians-Universität München}
  \city{\relax}
  \country{\relax}
}
\affiliation{
  \institution{Munich Center for Machine Learning}
  \city{\relax}
  \country{\relax}
}
\affiliation{
  \institution{Center for Digital Technology and Management}
  \city{Munich}
  \country{Germany}
}
\email{samuel.valenzuela@lmu.de}

\author{Johannes Kinder}
\affiliation{
  \institution{Ludwig-Maximilians-Universität München}
  \city{\relax}
  \country{\relax}
}
\affiliation{
  \institution{Munich Center for Machine Learning}
  \city{Munich}
  \country{Germany}
}
\email{johannes.kinder@lmu.de}

\copyrightyear{2026}
\acmYear{2026}
\setcopyright{cc}
\setcctype{by}
\acmConference[ICPC '26]{34th IEEE/ACM International Conference on Program Comprehension}{April 12--13, 2026}{Rio de Janeiro, Brazil}
\acmBooktitle{34th IEEE/ACM International Conference on Program Comprehension (ICPC '26), April 12--13, 2026, Rio de Janeiro, Brazil}
\acmPrice{}
\acmDOI{10.1145/3794763.3794795}
\acmISBN{979-8-4007-2482-4/2026/04}

\begin{document}

\begin{abstract}
Binary function embedding models are trained to encode the semantics of binary code in such a way that they can be generalized to a variety of reverse engineering tasks, such as binary code search, vulnerability detection, or malware classification. While many models only take the function in question as contextual input, there have been successful attempts to improve function embeddings by leveraging information from the call graph. In this study, we dissect the implications of these embedding refinements. We conduct experiments using a range of graph-based models on the embeddings generated by two state-of-the-art binary function embedding models. Integrating inter-procedural context, we show that improvements on binary code similarity detection (BCSD) will not necessarily generalize to downstream tasks, neither of semantic nor of syntactic nature. More generally, we find that optimizing for semantic similarity tasks correlates with worse performance on syntactic tasks. By conducting an explanatory analysis on the dataset, we find that the call graph-based enhancements significantly enhance the robustness of embeddings, particularly in scenarios where the initial models struggle. Furthermore, we observe that the added context is more beneficial for namespace-related functions than for those focused on individual logic, confirming that the call graph can be leveraged most effectively in context-dependent scenarios.
\end{abstract}

\begin{CCSXML}
<ccs2012>
   <concept>
       <concept_id>10002978.10003022.10003465</concept_id>
       <concept_desc>Security and privacy~Software reverse engineering</concept_desc>
       <concept_significance>500</concept_significance>
       </concept>
   <concept>
       <concept_id>10010147.10010257.10010293.10010294</concept_id>
       <concept_desc>Computing methodologies~Neural networks</concept_desc>
       <concept_significance>500</concept_significance>
       </concept>
 </ccs2012>
\end{CCSXML}

\ccsdesc[500]{Security and privacy~Software reverse engineering}
\ccsdesc[500]{Computing methodologies~Neural networks}

\keywords{Binary analysis, Binary code similarity detection, Call graphs, Explainability, Function embeddings, Graph neural networks}

\maketitle

\section{Introduction}

Binary code analysis is a subfield of program analysis dealing with situations in which the original source code of a compiled program is not available to the reverse engineers. Applications range from vulnerability detection~\cite{gao2018vulseeker,luo2023vulhawk} and malware classification~\cite{raff2018malware} to more general reverse engineering tasks such as binary code similarity detection (BCSD)~\cite{xu2017neural,liu2018alphadiff,massarelli2019safe,yu2020order,wang2022jtrans,wang2024clap} and function labeling~\cite{he2018debin,david2020neural,jin2022symlm,patrick2023xfl,benoit2025blens}. Additionally, syntactic binary analysis tasks such as compiler provenance recovery~\cite{rosenblum2011recovering}---i.e., identifying aspects including the compiler version or optimization level---help in understanding and working with the code~\cite{ji2021vestige}.

A particularly challenging aspect of low-level code comprehension is that two pieces of binary code can differ greatly in their structure and instruction sequences, yet implement the same functionality and thus be semantically equivalent. This also applies to binary programs compiled from the same source when using different compiler options or optimization levels. Crucially, binaries are typically stripped of debug information like variable names that may offer helpful clues regarding the code semantics.

In light of these intricacies, deep learning approaches have gained wide adoption for binary analysis tasks in recent years to help reverse engineers navigate machine code more effectively. As disassembled binary code can be read in a similar manner to sequential text, many approaches are inspired by advances in the NLP domain and make use of the way code is structured. For instance, distant instructions are oftentimes closely related through jumps or function calls. Different approaches have been adopted to model these relationships. While recent Transformer-based models capture control flow by sharing parameters between embeddings~\cite{wang2022jtrans,wang2024clap}, other methods employ graph-based models on different program representations such as the control or data flow graph~\cite{xu2017neural,gao2018vulseeker,yu2020order,luo2023vulhawk}.

\begin{figure*}[t]
    \centering
    \includegraphics{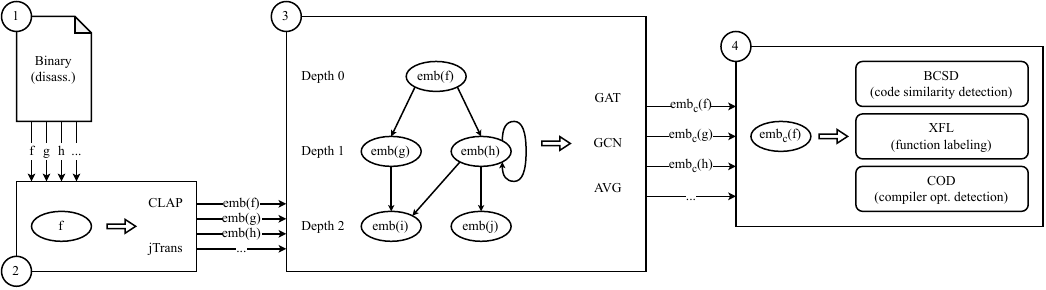}
    \caption{Overview of our workflow. First, a binary is disassembled and all functions are extracted. These are then processed by a binary function embedding model. The resulting embeddings for each function populate the nodes of the call graph and are updated using a graph-based model. The context-enhanced embeddings are then used as an alternative input to different machine learning tasks.}
    \Description{A flow chart illustrating a four-step methodology. Step one shows a disassembled binary providing individual functions such as f, g, and h as an output. Step two shows these functions being processed by a pretrained model such as CLAP or jTrans, generating an initial embedding for each function. Step three depicts a call graph where nodes are initialized with these embeddings. A node for the embedding of f at depth zero points to the nodes for the embeddings of g and h at depth one, indicating that f calls g and h. These nodes corresponding to g and h point to other functions at depth two, and h also points at itself. This graph is processed by a graph neural network or the averaging approach to produce an updated embedding for each function. Step four shows these updated embeddings being used as an input for code similarity detection, function labeling, and compiler optimization detection.}
    \label{fig:methodology}
\end{figure*}

Tuning graph nodes on their surrounding context has proven to be beneficial not only on an instruction level, but also on a function level~\cite{guo2022exploring}. Particularly given that the Transformer-based state of the art does not scale well to incorporate the entire program context, architectures have been emerging which extract information from the call graphs for different types of tasks including BCSD~\cite{liu2018alphadiff,guo2022exploring,yu2023cfg2vec,wang2025binenhance}, function labeling~\cite{jin2022symlm,yu2023cfg2vec}, and compiler provenance recovery~\cite{guo2022exploring,ji2021vestige}. Taking BCSD as an example, the intuition behind leveraging the call graphs to improve the quality of binary function embeddings is that the embeddings of a function $f$ and a function $g$ should approach each other if $g$ calls function $h$ and $f$ is semantically equivalent to $g$ but inlines $h$. Given that the average function in our dataset calls 4.4 distinct internal functions (i.e., excluding external library calls) the number of considered instructions is almost tripled from 84.3 to 239.0 in a single step and highlights why models can benefit from this additional context.

While the aforementioned approaches successfully incorporate call graphs into the model training for various tasks, little attention has been paid to the application of the resulting embeddings to downstream tasks without task-specific embedding refinements. More generally, there is no literature to our knowledge that systematically examines how such semantic tasks affect the model's retention of technical details and as such how generalizable the model is not only to other semantic tasks, but also to syntactic tasks such as compiler provenance recovery. Furthermore, this line of work has mainly shown \textit{that} the consideration of call graphs improves model performance, but there are no comprehensive studies on \textit{how} or for which types of functions this approach is especially fruitful.

In this paper, we explore how call graphs can be leveraged in the pretraining stage using BCSD. Inspired by related work~\cite{guo2022exploring,wang2025binenhance}, we use two variants of graph neural networks (GNNs) in order to leverage information from the call graph. The goal is to intervene in cases that previously viewed internal function calls as unknown \texttt{call} instructions, and to incorporate their function embeddings as meaningful semantic input for the embedding of the caller function. In our evaluation, we compare how the performance of both GNNs and a simple averaging baseline develops across varying degrees of inter-procedural context (call depths). Beyond the BCSD task, we extend the application of the context-enhanced embeddings to a semantic function labeling task as well as a syntactic compiler provenance task. Moreover, we conduct an explanatory analysis to uncover in which cases the embedding refinements prove to be most beneficial. For this, we split the dataset into various dimensions and compare how well the models perform for the different function groups. Thereby, we make the following contributions:
\begin{enumerate}
    \item We show that adding a pretraining task on the call graph can improve embeddings for BCSD, but worsen their generalizability to other semantic downstream tasks like function labeling.
    \item We show that training and better performance on semantic similarity tasks correlates with a decreased performance on syntactic tasks.
    \item We show that the contextual additions for BCSD make the embeddings more robust to larger context sizes, diversity within function pairs, and technical implementations.
    \item We identify that the enhanced embeddings improve more for namespace-related functions than for other functions.
\end{enumerate}
Our code and data is available online\footnote{https://github.com/lmu-plai/binary-callgraph-pretraining}.

\section{Methodology}
\label{sec:methodology}

\autoref{fig:methodology} provides a general overview of our workflow, which consists of four main steps:
\begin{enumerate}
    \item We disassemble an input binary and extract all functions as well as the call graph, a directed graph where each node corresponds to an internal function and each edge denotes a caller-callee relationship.
    \item We generate an embedding for each function using a pretrained model which we refer to as the backbone.
    \item After initializing its nodes with the corresponding functions' pretrained embeddings, the call graph is processed by a model which updates each node's embedding by aggregating its own features with those of its callees within a specified call depth.
    \item The resulting context-enhanced embeddings are then used for various function-level machine learning tasks as outlined in the following sections.
\end{enumerate}

We use IDA Pro~\cite{ida_pro} for disassembly and generate the initial function embeddings with CLAP~\cite{wang2024clap} and jTrans~\cite{wang2022jtrans} as backbones. Both are state-of-the-art Transformer-based models that also base their preprocessing pipeline on IDA Pro. Besides implementation-specific differences that provide CLAP with more information obtained by IDA Pro, the key difference lies in their training method. Initially, both models are trained via masked language modeling, a common pretraining task known from the NLP domain. However, after that, CLAP is trained using natural language supervision to align the binary code embeddings with explanations of the source code, whereas jTrans attempts to predict the targets of jump instructions.

For creating the context-enhanced function embeddings, we compare a simple averaging baseline and two prominent GNN variants: The Graph Convolutional Network (GCN)~\cite{kipf2017semi} and the Graph Attention Network (GAT)~\cite{velivckovic2017graph} are two GNNs that employ layer-specific weights, effectively distinguishing between different call depths. The model architectures differ in the way they aggregate information from neighbors. GCNs apply predefined weights at inference time, whereas GATs leverage an attention mechanism and thus can learn to evaluate the importance of neighbors---or in our case, callee functions. For each of these three model types, we consider a call depth from 1 to 5, resulting in 15 models per backbone. This means that, for a call depth of $n$, we extract a subgraph containing $n$ steps of node successors to the function in question, and train a GNN with $n$ layers.

All of our experiments are conducted on a dataset derived from BinaryCorp-26M~\cite{wang2022jtrans}, a large set of binaries compiled with different optimization settings which both CLAP and jTrans were trained on. The complete dataset of binaries, including all programs and functions referenced later in this paper, is available online\footnote{https://github.com/vul337/jTrans}. After stripping each binary to remove all debugging symbols, we disassemble and extract all functions. Next, we apply two filtering steps. First, we only consider functions with at least four instructions to reduce the noise caused by trivial functions such as empty functions, basic wrappers, or compiler-generated PLT stubs. To facilitate the training process described below, we only keep one function pair per source function if compilation with different optimization configurations results in binary code with differing opcode hashes. This way, we avoid function pairs that only differ in register allocations or the binary's address layouts. The resulting dataset consists of approximately 1.4M function pairs from 34.3k binaries compiled from 9.4k source programs. We retain the train-test split of BinaryCorp-26M and finetune hyperparameters on a secluded part of the train set.

\section{Tasks}

We train the GNNs on the BCSD task, ensuring a comparable approach to related work on cross-configuration binary code detection by \citeauthor{guo2022exploring}~\cite{guo2022exploring}. After training, we evaluate the generalizability of the refined function embeddings via two downstream tasks, namely eXtreme Function Labeling (XFL)~\cite{patrick2023xfl} and Compiler Optimization Detection (COD).

\subsection{Binary Code Similarity Detection}

In BCSD, the goal is to measure the similarity between two instances of binary code. All GNNs are trained on BCSD using the contrastive InfoNCE loss~\cite{oord2018representation}, specifically the NT-Xent variant~\cite{chen2020simple} as implemented by \citeauthor{silva2020exploringsimclr}~\cite{silva2020exploringsimclr}, which rewards the similarity of functions compiled from the same source function and penalizes the similarity between other function pairs. Per batch of positive function pairs, the remaining functions in the batch are used as negative samples for each function. Explored hyperparameters include the batch size, the dropout rate, the learning rate and weight decay for the Adam optimizer, as well as the temperature for the InfoNCE loss. Starting with common values found in literature and conducting empirical evaluations to decide on fixed values for our experiments with both GNNs, each model is then trained for 50 epochs using the largest possible batch size of 8,192 function pairs on our system with NVIDIA H100 GPUs.

To sanity-check the implementation of our approach, we conduct an additional experiment where each function is assigned a random function embedding. In this scenario, we ensure that the GNN's performance does not improve significantly after training, as that would reveal a flaw in the methodology.

\subsection{Function Labeling}

The first downstream task we consider is XFL, a recent approach to predict function names which leverages the output of pretrained embedding models in a straightforward manner. We select this task for our evaluation because profound semantic understanding of the binary function is required to make accurate predictions. In essence, XFL creates a label space for a dataset and decomposes function names into a set of normalized labels. These are then predicted in a multilabel classification task by passing the precomputed embeddings through a tree-based classifier. This methodological split between the function embedding and the function label prediction stage makes it difficult to finetune deep learning-based embedding models on this task. As a result, it makes sense to consider alternative methods to refine the embeddings, for instance on the call graph.

Adapting it to our dataset and embeddings, we reuse the original XFL implementation as far as possible and leave functionality such as the handling of common statically identifiable function names untouched. As our main goal is to compare the generalization capabilities of precomputed embeddings rather than achieving a new state of the art, we abstain from any XFL-specific hyperparameter tuning and use the default values set by the authors. For this task, we consider only C functions and exclude C++ functions due to their mangled function names, resulting in a subset of approximately 351.3k functions.

\subsection{Compiler Optimization Detection}

To examine the retention of technical details in the embeddings, we additionally perform a subtask of compiler provenance recovery. Leveraging the metadata in the BinaryCorp dataset, we introduce COD as a second downstream task. Predicting the underlying compiler optimization allows us to evaluate how integrating information from a function's callees affects the resulting embeddings' sensitivity to low-level structural variations induced by the compiler. Moreover, by training the GNN models on BCSD, we assess how the focus on higher-level semantic similarity tasks affects the downstream performance on syntactic tasks.

\begin{figure*}[t]
    \centering
    \includegraphics{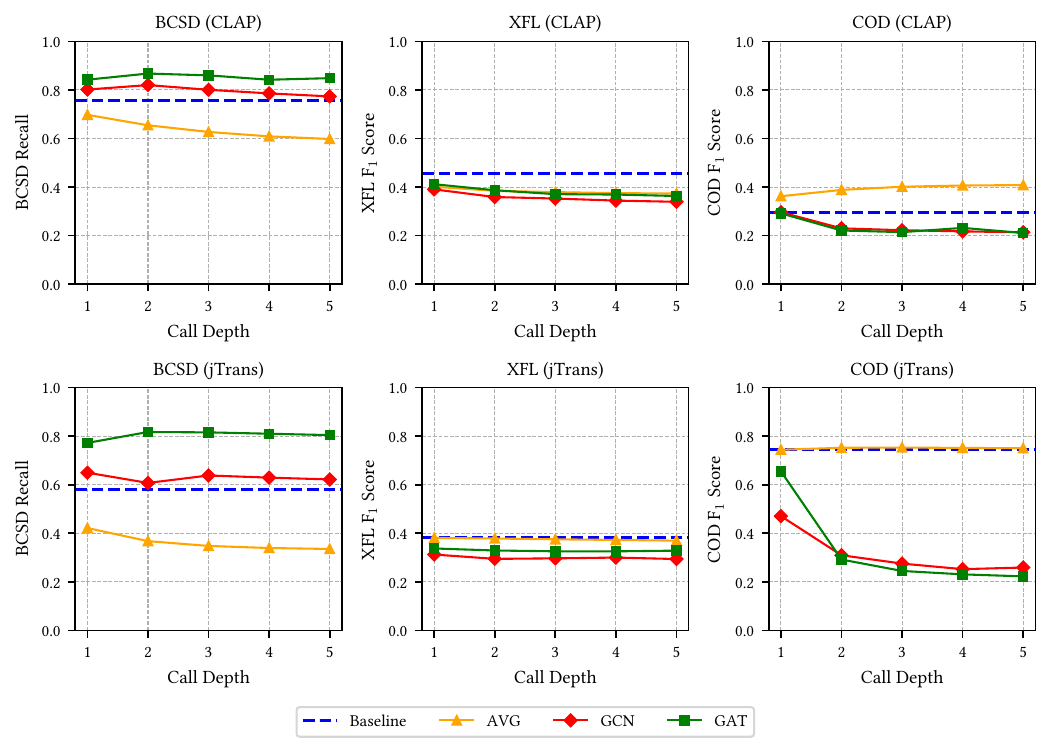}
    \caption{Task-specific results of CLAP and jTrans (Baseline) as well as the GNNs (GCN and GAT) and averaging approach (AVG) given different depths of the call graph. The diagrams plot the recall@1 scores for binary code similarity detection (BCSD), micro F$_\mathbf{1}$ scores for the function labeling task (XFL), and macro F$_\mathbf{1}$ scores for the compiler optimization detection task (COD).}
    \Description{Six line charts in a two-by-three grid. The top row shows the CLAP-backed results; the bottom row shows results with the jTrans backbone. The three columns correspond to the tasks: binary code similarity detection, function labeling, and compiler optimization detection. Each chart plots task performance on the y-axis against call depth from one to five on the x-axis for the Graph Attention Network, Graph Convolutional Network, and averaging approach. Baseline scores for CLAP and jTrans are also shown. CLAP generally outperforms jTrans except in compiler optimization detection. Graph neural networks improve performance for binary code similarity detection, whereas the averaging approach performs best for compiler optimization detection. For function labeling, all context-enhanced models perform worse than the baselines.}
    \label{fig:task_results}
\end{figure*}
As COD is a multiclass classification task that labels the functions with a compiler optimization from \{\texttt{O0}, \texttt{O1}, \texttt{O2}, \texttt{O3}, \texttt{Os}\}, a linear layer is trained on the precomputed embeddings to predict the correct optimization level. Similar to the setup used for XFL, our approach is equivalent to freezing all parameters of both CLAP and jTrans as well as the GNNs for the downstream task. Training of the linear head is performed by minimizing the cross-entropy loss. Initially, we also experimented with multilayer perceptrons (MLPs) containing one hidden layer, though the added complexity did not yield significantly different results. By selecting one function per function pair, our dataset consists of approximately 1.4 million functions and the model is trained for 50 epochs with a batch size of 32.

\section{Results}
\label{sec:results}

In this section, we evaluate all pretrained baseline models and graph-based models on the three tasks outlined in \autoref{sec:methodology}. \autoref{fig:task_results} plots the test set performance of each model per task. The top row depicts the diagrams using CLAP as a backbone, whereas the bottom row concerns the jTrans backbone. The following sections provide an overview of the evaluation method and an analysis of the results. As we regard the call graph-based models with a call depth from 1 to 5, we refer to them as AVG($n$), GCN($n$), and GAT($n$) for a call depth of $n$, respectively.

\subsection{Embedding Refinement on BCSD}
\label{sec:results_bcsd}

To evaluate the BCSD task, we use a pool size of 1,000 function pairs (2,000 functions in total). For each function, this pool consists of the single positive counterpart from the same source and 1,998 negative function samples drawn from the other 999 pairs. The first column in \autoref{fig:task_results} shows the recall@1 scores of each model after training. This metric represents the share of functions for which the correct positive counterpart is the top-ranked result, based on cosine similarity, among all 1,999 candidate functions. Following the approach of CLAP~\cite{wang2024clap} and jTrans~\cite{wang2022jtrans}, the mean reciprocal rank (MRR) was also taken into consideration as a performance metric. This metric proportionally accounts for the rank of the correct item, with higher scores for items ranked closer to the top. However, the observable patterns do not differ considerably between both metrics. Due to its more direct interpretability, we choose to only focus on the recall@1 metric.

Multiple insights can be obtained from these results. Firstly, we are able to reproduce previous findings that CLAP performs significantly better on the BCSD task than jTrans~\cite{dannehl2025instructions,wang2024clap}. Moreover, it becomes evident that GNN-based models are able to extract meaningful information solely from the call graph in order to improve a backbone's performance on detecting same-source functions, confirming the intuition applied in previous work~\cite{guo2022exploring,wang2025binenhance}. With exception of the generally worst-performing model---the GCN on jTrans---the call depth of 2 appears to strike the best balance between contextual yet relevant information on our dataset. This appears to be slightly more shallow than the typical call depth of 4 reported in related work~\cite{ji2021vestige,guo2022exploring,yu2023cfg2vec,wang2025binenhance}, highlighting that the optimal call depth is sensitive to the general experiment setup. In our experiments, the more sophisticated GNN variant clearly outperforms all other models. The usage of GAT(2) leads to a performance improvement of almost +0.11 from 0.758 to 0.867 for CLAP and an improvement of +0.24 from 0.579 to 0.817 for jTrans, almost achieving comparable results to the CLAP-based GATs.

While similar trends can be observed for both models, the variance between context-enhanced embeddings is much larger with the jTrans backbone than CLAP. We hypothesize that particularly the GATs learn to recognize similar call graph structures and callee embeddings and investigate this further in \autoref{sec:explanatory_analysis}. Taking into account the fact that the preprocessing pipeline CLAP retains more contextual information---though mostly about called library functions rather than internal calls---we believe the context-enhanced embeddings are able to address some of the shortcomings of jTrans compared to CLAP. Having said that, the results of the AVG models suggest that the new contextual information must be combined in a more involved manner, as none of the models outperforms the baseline set by the backbone, and the performance deteriorates further for larger call depths.

\subsection{Semantic Downstream Performance on XFL}

The middle column of \autoref{fig:task_results} presents the F$_1$ scores of each model on the XFL task. With a label space of approximately 1,024 distinct classes, the same class imbalance occurs on our dataset as discussed in the original paper~\cite{patrick2023xfl}. Therefore, we follow the same approach and only investigate the micro-averaged scores, i.e., where all true positives, false positives, and false negatives are aggregated across classes rather than weighting each class equally.

The most striking insight is that both CLAP and jTrans outperform every model that attempts to enhance the embeddings with contextual information from the call graph. It appears that the semantic aggregation of functions learned on BCSD is not directly applicable to XFL. The comparison between the graph-based models underscores this further. While the simple averaging of embeddings in the call graph is significantly outperformed by the GNNs on the BCSD task, the AVG models perform comparably, if not even better, than the GNNs on the XFL task.

These results suggest that same-source binary code detection may not be a suitable task to ensure a semantic understanding that can be generalized to other tasks. At the same time, it is not possible to conclude that function embeddings cannot be refined on the call graph to improve for function labeling tasks. In \autoref{sec:namespace_labels}, we investigate more closely in what way the investigated models may have benefited from the training regardless.

\subsection{Impact of Semantic Similarity Pretraining on Syntactic Downstream Tasks}
\label{sec:results_cod}

The right column of \autoref{fig:task_results} shows the models' F$_1$ scores on COD. As there is a minor class imbalance caused by the dataset which should be evened out, we use the macro-averaged F$_1$ score as opposed to micro-averaging as done for XFL. The patterns visible in the results of this syntactic task contrast strongly with the more semantic BCSD task. Firstly, neither GNN model type benefits the performance after being trained on refining the embeddings for BCSD. Instead, the performance keeps deteriorating particularly starting at a call depth of 2. The AVG models, on the other hand, outperform their backbone and continue improving with access to larger call depths when using CLAP as a backbone. These observations suggest that, to a certain extent, the simple approach of averaging all embeddings aggregates low-level technical details and patterns much better than high-level semantic information.

That being said, it is important to note that neither CLAP nor jTrans reach evaluation scores similar to those reported in literature that investigates task-specific models which are specifically developed for compiler provenance recovery~\cite{rosenblum2011recovering,he2022binprov}. While it is intuitive that models trained to abstract away low-level features will perform worse on a syntactic task like COD, it is remarkable how much the performance differs between both backbones. CLAP clearly outperforms jTrans on both other tasks, whereas it is significantly surpassed on COD with an F$_1$ score of 0.295 as opposed to 0.743. In this case, the reduced access to information reconstructed by IDA in the preprocessed inputs may actually benefit jTrans on the COD task. This establishes a consistent pattern across both the pretrained backbones and the refined call graph-based models, highlighting the tension between abstractive power and technical precision and thus an inverse correlation where stronger semantic similarity capabilities come at the cost of lower performance on syntactic downstream tasks such as COD.

\section{Explanatory Analysis}
\label{sec:explanatory_analysis}

In this section, we dismantle the semantic task results from \autoref{sec:results} by grouping them across various dimensions, including function and call graph size, compiler optimization, and source language. Using the normalized labels from XFL, we also inspect the results per label, categorizing them into namespace-related and other labels as described in \autoref{sec:namespace_labels}. Specifically targeting BCSD, we additionally group by function pair properties such as the similarity between both functions, the highest similarity between any of their callees, and whether their call graphs are likely isomorphic or not. Our analyses focus on answering the following four research questions:

\begin{itemize}
    \item \textbf{RQ1:} How beneficial is the embedding refinement on the call graph for semantic tasks if large amounts of contextual information is available?
    \item \textbf{RQ2:} How robust are the embedding refinements toward diversity within function pairs?
    \item \textbf{RQ3:} How robust are the embedding refinements toward different technical implementations?
    \item \textbf{RQ4:} Are the call graph-based models more advantageous on certain types of function labels?
\end{itemize}

In our deep-dive, we focus on the best-scoring instance of each model family for the respective task. As such, we neglect the GCN models entirely due to their inferior performance compared to the GATs. Additionally, we only take into account grouped results with a sample size of at least 200. While we take both CLAP- and jTrans-based results into consideration, the data points we present focus mostly on CLAP as a backbone due to its superior performance. Because CLAP embeddings provide a more meaningful starting point of the node embeddings, we expect the insights to be more informative. However, we do point out cases if the patterns differ for jTrans.

\subsection{Benefit of More Context}

Intuitively, the contextual embedding refinements should be particularly useful when more context is available. Take, for instance, the \texttt{CB\_InputChanged} function from \texttt{photivo-git-photivo} on the BCSD task. Both function instances compiled with the optimizations \texttt{O1} and \texttt{O3} call nearly 140 distinct internal functions. Their CLAP embeddings do not show any significant similarity, with both functions having a higher similarity to more than a quarter of the remaining batch. However, using both the GAT(2) model and even AVG(1), the function pair is correctly matched among all 1,000 function pairs.

To quantitatively verify our intuition regarding the impact of context size on the performance of call graph-enhanced function embeddings, we consider both the number of nodes in the call graph as well as the number of instructions in the regarded function itself. Specifically, we group the functions logarithmically in buckets labeled $\lfloor\log_2(\mathrm{Nodes})\rfloor+1$ and $\lfloor\log_2(\mathrm{Instructions})\rfloor+1$, representing the number of bits needed to store the number of nodes or instructions, respectively. For the sake of comparability between models, we only present the number of nodes for a depth of 1. This equates the number of distinct internal callees plus one, as the key function in question is also included in the call graph.

Starting with the XFL task, all models demonstrate better performance for the shortest and longest functions with the fewest and the most callees. On the one hand, this means that all models profit from large amounts of available context. The reasoning for the performance increase given short functions is addressed further in \autoref{sec:namespace_labels}. Regardless, the trend is comparable across all models including the backbones. That is, the performance change on XFL after the training on BCSD with the call graph does not depend noticeably on the function size.

\begin{figure}[t]
    \centering
    \includegraphics{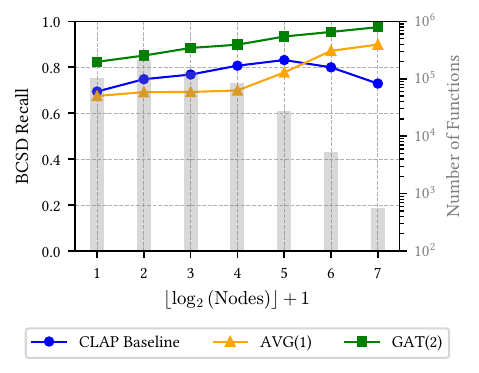}
    \caption{Grouped BCSD results of CLAP-backed models by exponential number of nodes (i.e., number of bits needed to represent the number) in the call graph (call depth 1).}
    \Description{A combined line and bar chart with two y-axes. The x-axis shows the logarithm base two of the number of nodes in the call graph with depth one, rounded down, plus one. The left y-axis represents the recall scores for binary code similarity detection. The scores keep increasing for the averaging approach on a call depth of one and the Graph Attention Network on a call depth of two. For the CLAP baseline, the scores start decreasing again for the highest numbers of nodes. The right y-axis displays the number of functions behind each data point on a logarithmic scale from 100 to 1 million. The bar chart peaks at the value 2 and decreases steadily afterward.}
    \label{fig:grouped_results_bcsd_nodes}
\end{figure}

In contrast, the embedding refinement does show a noticeable improvement on the BCSD task particularly for large functions and call graphs. Exemplarily, \autoref{fig:grouped_results_bcsd_nodes} plots the CLAP-backed model performance based on the number of nodes in the call graph and displays the number of functions per bucket. While the call graph-enhanced embeddings demonstrate a steady increase in performance with more unique callees, CLAP performance rapidly declines starting at a number of $2^{(5+1)}=64$ nodes in the call graph.

Intuitively, both grouping dimensions---number of call graph nodes and number of instructions comprising the function---are correlated with each other, as larger functions will have more call instructions with more distinct callees on average. Therefore, it comes as no surprise that largely equivalent patterns emerge when focusing on the instruction buckets. It is important to note that grouping by both dimensions jointly yields comparable results as well---the only exception being jTrans, which performs best on longer functions with smaller call graphs, likely due to the fact that its preprocessing pipeline obscures potentially helpful contextual information. Particularly for the best-performing GAT, on the other hand, the performance generally increases both with more callees and with more instructions.

The key explanation for the performance drop of CLAP and jTrans lies in their limited context size of 1,024 tokens and 512 tokens, respectively. Taking into account that instructions typically consist of several tokens, this limit coincides well with the cutoff point before the performance drop. As the graph-based models examined in this paper do not aggregate semantic information on a token- or instruction-level, but rather on a function-level, they are able to effectively work against the fixed-window-size restriction of Transformer-based function embedding models. Furthermore, in the case of jTrans, the GAT(2) model is able to beneficially aggregate the callee embeddings to counteract the backbone's performance decrease given larger call graphs.

\textbf{Summary:} \textit{Contextual models learn to harness larger amounts of context for BCSD, whereas the performance of CLAP and jTrans decreases eventually due to the limited size of their context window.}

\subsection{Semantic Robustness on BCSD}

In this section, we analyze how similarities and divergences between paired functions impact the models' performance on BCSD. The function \texttt{window\_editor\_map\_draw\_panels} from the program \texttt{augustus-game-git-augustus-game} serves as a motivating sample. The function pair using the optimizations \texttt{O0} and \texttt{O2} yields CLAP embeddings with a cosine similarity close to 0, and there are no remotely identical function embeddings with a similarity above 0.75 in their differently sized call graphs with a depth of 2. Nonetheless, both GAT(2) and AVG(1) successfully aggregate the embeddings such that the function pair is correctly matched in its evaluation pool, raising the question how these models deal with cases in which the function embeddings or call graphs do not display significant similarities.

\begin{figure}[t]
    \centering
    \includegraphics{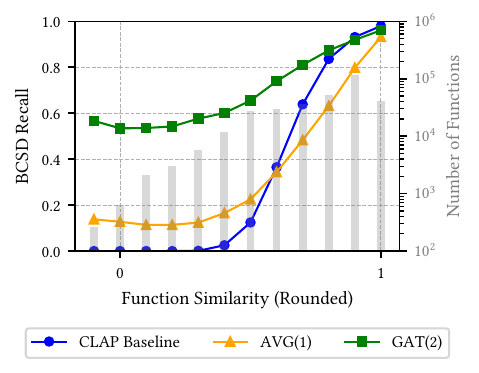}
    \caption{Grouped BCSD results of CLAP-backed models by function similarity.}
    \Description{A combined line and bar chart with two y-axes. The x-axis shows the cosine similarity between both functions, rounded to 0.1 steps. The left y-axis represents the recall scores for binary code similarity detection. The scores for the CLAP baseline are highest for the largest function similarities and drop the quickest, crossing the line of the Graph Attention Network on a call depth of two between a cosine similarity of 0.8 and 0.9, and the line of the averaging approach on a call depth of one between 0.5 and 0.6. The right y-axis displays the number of functions behind each data point on a logarithmic scale from 100 to 1 million. The bar chart peaks at the value 0.9 and decreases steadily for lower similarities.}
    \label{fig:grouped_results_bcsd_func_similarity}
\end{figure}

To attain a quantitative conclusion, we examine the grouped results for each related dimension. As the patterns align for both backbones, we only present results for CLAP. \autoref{fig:grouped_results_bcsd_func_similarity} plots the recall@1 score grouped by the rounded cosine similarity between both functions' CLAP embeddings. Note that cosine similarities can assume values within $[-1,1]$, which is why the data points also include negative similarities. It becomes visible that for pairs with a similarity near 1, it may not be beneficial to add more context from the call graph. However, as anticipated, particularly the CLAP baseline's model decreases rapidly toward 0 for lower similarities. In contrast, the GAT(2) model always demonstrates a performance above 0.53 for all buckets, showcasing that the GNN is capable of effectively integrating the knowledge from the call graph in more than half of the cases where the function bodies themselves are entirely dissimilar.

\begin{figure}[t]
    \centering
    \includegraphics{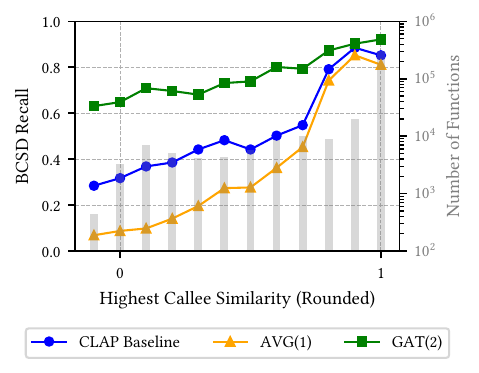}
    \caption{Grouped BCSD results of CLAP-backed models by highest similarity between callees (call depth 1).}
    \Description{A combined line and bar chart with two y-axes. The x-axis shows the highest cosine similarity between two callees in both function's call graphs, rounded to 0.1 steps. The left y-axis represents the recall scores for binary code similarity detection. All models achieve higher scores for higher similarities. The Graph Attention Network on a call depth of two always outperforms the CLAP baseline, particularly for similarities below 0.8. The CLAP baseline always outperforms the averaging approach on a call depth of one. The right y-axis displays the number of functions behind each data point on a logarithmic scale from 100 to 1 million. The bar chart peaks at the value 1.0 and displays a downward trend for smaller callee similarities.}
    \label{fig:grouped_results_bcsd_callee_similarity}
\end{figure}

One natural reason behind this performance is that some function pairs call other functions which do not differ across compiler optimizations and therefore support the graph-based plug-in models to bring the function embeddings closer together. However, the data shows that GAT(2) also yields superior results in more intricate scenarios in which none of the callees are similar between both functions in question. This is shown in \autoref{fig:grouped_results_bcsd_callee_similarity}, which plots the recall@1 score grouped by the highest similarity between any two callees in the function pair's call graphs. While this diagram only refers to the functions' direct neighbors in the call graph, it should be noted that the trends stay the same with a call depth of 2. Drawing from these insights, we find that the GNN is capable of meaningfully aggregating all embeddings in the call graph despite the individual embeddings yielding low similarity scores to their counterparts in the function pair.

In addition to the node embeddings, similar insights can be made with regard to the graph structure. \autoref{tab:grouped_results_bcsd_is_likely_isomorphic} lists the models' performance for cases where both call graphs of depth 2 are non-isomorphic or likely isomorphic, as determined using the NetworkX library~\cite{hagberg2008networkx}. On this dimension, particularly the jTrans-based results reinforce our previous findings on the semantic robustness of GAT(2) on BCSD, displaying a difference in recall@1 scores of only 0.06 compared to the baseline's 0.20.

\begin{table}[t]
\centering
\caption{Grouped BCSD recall@1 results by graph isomorphism of function pair's call graphs (call depth 2).}
\label{tab:grouped_results_bcsd_is_likely_isomorphic}
\begin{tabular}{lrr}
\toprule
& \textbf{non-isomorphic} & \textbf{likely isomorphic} \\
\midrule
\textbf{Functions} & 385,218 & 173,506 \\
\midrule
\textbf{CLAP Baseline} & 0.702 & 0.878 \\
\textbf{CLAP AVG(1)} & 0.613 & 0.881 \\
\textbf{CLAP GAT(2)} & 0.836 & 0.933 \\
\textbf{jTrans Baseline} & 0.517 & 0.714 \\
\textbf{jTrans AVG(1)} & 0.277 & 0.741 \\
\textbf{jTrans GAT(2)} & 0.796 & 0.860 \\
\bottomrule
\end{tabular}
\end{table}

\textbf{Summary:} \textit{The GAT(2) model demonstrates more robust BCSD performance in cases where the function embeddings or the call graphs are less similar.}

\subsection{Technical Robustness on BCSD}

Turning to the technical robustness of the regarded models, an expressive sample is the C++ function \texttt{decompress\_etc\_eac} from the program \texttt{nvidia-texture-tools-git-nvcompress} compiled with the optimizations \texttt{O0} and \texttt{Os}---the pair on which CLAP demonstrates the worst performance. In this case, the CLAP embeddings are more similar to more than a third of the remaining batch than to each other, while being matched correctly by GAT(2) and AVG(1), suggesting that embedding refinements become more robust to technical details.

To evaluate this robustness more systematically, we first consider the source language, i.e., if the binary code originates from a C or a C++ function. \autoref{tab:grouped_results_bcsd_source_language} shows the recall@1 values for all models, revealing the same patterns for CLAP- and jTrans-backed models. Without falling behind on C functions, the GAT(2) model achieves much more similar performance on C++ functions, with a performance difference of less than 0.08 compared to the twice as large drop by the CLAP baseline. Investigating possible correlations between the source language and other examined dimensions, we find that there is no integral correlation between C++ functions and a larger graph size. However, the stronger level of abstraction found in C++ compared to C can be unraveled by splitting the grouped results into non-isomorphic and likely isomorphic graphs. Here it becomes apparent that many more graphs are isomorphic for C functions than for C++ functions. Nonetheless, filtering only for non-isomorphic graph pairs and thus subtracting out correlations between source languages and the share of isomorphic graph pairs, the GAT(2) model continues to demonstrate a much smaller performance drop from 0.910 to 0.825 from C to C++ functions, whereas the scores for CLAP drop from 0.875 to 0.676.

\begin{table}[t]
\centering
\caption{Grouped BCSD recall@1 results by source language.}
\label{tab:grouped_results_bcsd_source_language}
\begin{tabular}{lrr}
\toprule
& \textbf{C} & \textbf{C++} \\
\midrule
\textbf{Functions} & 108,616 & 450,108 \\
\midrule
\textbf{CLAP Baseline}   & 0.889 & 0.725 \\
\textbf{CLAP AVG(1)}     & 0.890 & 0.650 \\
\textbf{CLAP GAT(2)}     & 0.930 & 0.851 \\
\textbf{jTrans Baseline} & 0.780 & 0.529 \\
\textbf{jTrans AVG(1)}   & 0.692 & 0.356 \\
\textbf{jTrans GAT(2)}   & 0.884 & 0.799 \\
\bottomrule
\end{tabular}
\end{table}

That being said, an indicator for the technical robustness of the GAT-based model that has less potential for correlations with other factors is the differentiation of results between compiler optimizations. For this, we group the BCSD results by pair of compiler optimizations such as \texttt{O0,O1} or \texttt{O1,Os}, and regard the absolute range as well as the standard deviation between the different pairs. As can be seen in \autoref{tab:grouped_statistics_bcsd_compileopt}, the GAT(2) model demonstrates a significantly lower performance fluctuation across compiler optimization pairs compared to both CLAP and jTrans. For instance, while recall@1 scores differ by up to 0.23 for CLAP and 0.52 for jTrans, this difference is reduced to 0.13 and 0.18 using the GNN, respectively. While it is arguable that in the case of CLAP, both models triple their error rate on the worst-performing optimization pair compared to the best-performing one, the GAT(2) model demonstrates superior relative robustness using the jTrans backbone. Here, the error rate of jTrans increases almost fivefold, whereas GAT(2) remains at a threefold increase. However, in the absolute sense as argued above, we consider the GNN-based model more robust compared to CLAP as well. On BCSD, the call graph-based GNNs thus prove that their focus on aggregating higher-level information makes them less sensitive toward lower-level changes induced by different compiler options. Due to the limited transferability of these learnings to XFL, it is important to note that no such differences in robustness toward compiler optimizations can be observed for XFL. At the same time, with CLAP's XFL F$_1$ scores ranging from 0.406 to 0.457, for instance, the baseline models themselves do not demonstrate nearly as large susceptibility to such technical factors.

\begin{table}[t]
\centering
\caption{Statistics of BCSD recall@1 results grouped by compiler optimization pairs.}
\label{tab:grouped_statistics_bcsd_compileopt}
\begin{tabular}{lrrr}
\toprule
\textbf{Model} & \textbf{min} & \textbf{max} & \textbf{std} \\
\midrule
\textbf{CLAP Baseline}   & 0.650 & 0.884 & 0.100 \\
\textbf{CLAP AVG(1)}     & 0.561 & 0.878 & 0.131 \\
\textbf{CLAP GAT(2)}     & 0.802 & 0.933 & 0.048 \\
\textbf{jTrans Baseline} & 0.341 & 0.864 & 0.172 \\
\textbf{jTrans AVG(1)}   & 0.206 & 0.776 & 0.185 \\
\textbf{jTrans GAT(2)}   & 0.732 & 0.912 & 0.064 \\
\bottomrule
\end{tabular}
\end{table}

\textbf{Summary:} \textit{The GAT(2) model demonstrates more robust BCSD performance toward different source languages and compiler optimizations.}

\subsection{Impact of Label Classes}
\label{sec:namespace_labels}

Lastly, we aim to determine at a higher level which types of functions may benefit more from the application of the call graph-based models. For this, we analyze the performance change concerning both BCSD and XFL for each function label generated by XFL. As the number of labels is too large to systematically draw insights from the results, we classify them into namespace-related labels versus other types of labels, effectively distinguishing between labels outlining the function context and function-specific descriptors. As it is not feasible for us to perform a per-function determination of which label instances act as a namespace and which act as the actual function description, we assign all instances of each label to either the namespace class or the other class. For this, we prompted three different LLMs, namely Gemini 2.5 Pro~\cite{comanici2025gemini}, Claude Sonnet 4~\cite{claude_sonnet_4_news}, and DeepSeek R1 0528~\cite{deep_seek_r1_0528}. We adopted unanimous classifications after a brief manual verification, while resolving all disagreements through domain knowledge and online searches to identify library namespaces.

\begin{table}[t]
\centering
\caption{Grouped BCSD recall@1 and XFL micro F$_1$ results of CLAP-backed models by function label class. The best-scoring instance per model family is chosen for each task (AVG(1) and GAT(2) for BCSD, AVG(1) and GAT(1) for XFL).}
\label{tab:grouped_results_namespace_labels}
\begin{tabular}{lrr|rr}
\toprule
& \multicolumn{2}{c|}{\textbf{BCSD (recall@1)}} & \multicolumn{2}{c}{\textbf{XFL (micro F$_1$)}} \\
& \textbf{namespace} & \textbf{other} & \textbf{namespace} & \textbf{other} \\
\midrule
\textbf{Functions} & 55,134 & 278,614 & 27,567 & 139,307 \\
\midrule
\textbf{CLAP}      & 0.802 & 0.867 & 0.625 & 0.411 \\
\textbf{CLAP AVG}    & 0.799 & 0.868 & 0.607 & 0.342 \\
\textbf{CLAP GAT}    & 0.894 & 0.919 & 0.604 & 0.358 \\
\bottomrule
\end{tabular}
\end{table}

\autoref{tab:grouped_results_namespace_labels} presents the resulting BCSD and XFL scores grouped by the function label class for CLAP-backed models. On average for BCSD, functions with a stronger focus on the namespace perform worse. However, the margin is much smaller for the GAT(2) model, where the label class-specific scores are 0.025 apart from each other, compared to the baseline model with a margin almost three times the size. For XFL, in turn, all models perform significantly better on predicting the contextual labels compared to function-specific labels. While the GAT(1) model performs worse on the function labeling task than its baseline, as discussed in \autoref{sec:results}, the data shows that the performance drop is much less significant for namespace-related labels. While the F$_1$ score decreases by approximately 0.02 in this case, the score specific to other function labels drops by more than 0.05. This reaffirms our previous findings that the call graph-based models are more beneficially used in context-dependent scenarios such as predicting namespaces compared to cases that mostly concern a function individually.

Take, for instance, the following two exemplary labels providing an illustrative view on the effect of using the GNN model: \texttt{glfw}, the namespace for an OpenGL library and the 53rd most common label in our dataset of C functions, has a comparatively high F$_1$ score of 0.836 using CLAP, which is nevertheless slightly outperformed by the GAT(1) model which achieves an F$_1$ score of 0.844. In contrast, the most common label, \texttt{get}, though achieving a comparatively high score of 0.640 using CLAP, is not predicted as well by the GAT(1) model with an F$_1$ score of 0.533.

An additional noteworthy insight can be made by grouping jointly by the label class as well as the number of callees. Interestingly, the worst-performing group on BCSD and by far the best-performing group on XFL correspond to namespace-related function names without any internal calls. While a possible explanation on CLAP's side may be that these functions likely call external library functions particularly often---for instance, functions in the \texttt{java} namespace calling external functions with \texttt{java} in their name---this reasoning cannot be applied to jTrans. Here, such function calls are preprocessed as \texttt{call callfunc\_xxx}, shifting the explanation toward structural similarities between namespace labels. However, as this trend stays consistent throughout both the backbones as well as the embedding refinement models, we deem a thorough explanatory analysis at this point beyond the scope of this paper and leave it open for future work.

\textbf{Summary:} \textit{Performance changes of contextual models on semantic tasks are favorable for namespace-related functions compared to those focused on individual logic.}

\section{Threats to Validity}

\paragraph{Internal validity}

The threats to internal validity arise primarily from design choices made in this study. We focused on examining trends in model performance changes as opposed to achieving the highest possible metric scores. As such, we opted for widely known and comparatively simple GNNs instead of the latest state of the art, and did not perform XFL-specific hyperparameter tuning. Nonetheless, we acknowledge that our findings could partly be influenced by our experiment setup, as seen by the deviation of the optimal call depth in our experiments from previous literature referenced in \autoref{sec:results_bcsd}. Similarly, our contextual call graph-based models were trained only on BCSD. While specifically targeting downstream tasks that may not allow for simple finetuning of the embeddings, it must be noted that the selected training task may have restricted which information the models learned to capture. Finally, our two-step approach using a plug-in architecture potentially affected the interaction between intra- and inter-function information. Subject to computational constraints, architectures integrating contextual information from the call graph more directly into the encoder's training process could yield different patterns and would be an interesting avenue for future work.

\paragraph{External validity}

A main threat to external validity is the selection of our downstream tasks. In particular XFL's decoupled approach to predicting function names may limit the extent to which our insights generalize to related function labeling tasks such as BLens~\cite{benoit2025blens}, which optimizes the encoder's weights throughout the entire training process. Another external validity concern is the selected dataset. As all experiments are grounded on BinaryCorp, the observed behaviors might be influenced by the dataset's code distribution, compilers, or labeling schemes, and may not hold for other binary corpora. Moreover, the selected corpus is limited to benign binaries. Other trends may be observed when analyzing the context-enhanced embeddings on malware, particularly obfuscated code with complex function relationships, which may limit the extent to which our findings transfer to the software security domain.

\paragraph{Construct validity}

Our explainability analyses pose potential threats to construct validity. Although we explored a diverse and as comprehensive set of dimensions as possible, additional grouping factors may exist that were not captured in this study. Particularly higher-level semantic dimensions may yield more findings in addition to our analysis where we split by namespace versus function-specific labels. Similarly, we relied on a specific type of explainability approach centered on dataset slicing. Other established approaches, such as attribution techniques that identify the most salient parts of the input, have started being applied in the domain of binary code~\cite{dannehl2025instructions} and could provide complementary insights into how models leverage the call graph if applied systematically. Lastly, because our explainability analyses were performed on GNNs trained on the BCSD task, the patterns observed may be influenced by task-specific factors, and could differ if the models were trained on other tasks.

\section{Related Work}

A lot of recent literature is dedicated to creating generalizable embeddings of binary code semantics. BCSD is commonly the central task evaluated in this context. Well-known work in this realm includes models such as Asm2Vec~\cite{ding2019asm2vec} and SAFE~\cite{massarelli2019safe}, two approaches based on static embedding models like word2vec~\cite{mikolov2013efficient} that aggregate fixed token vectors to a function embedding. More recent models leverage the success of Transformers~\cite{vaswani2017attention} in the NLP domain and apply them to binary functions as well, thus tuning the embeddings more profoundly on the context of instructions. Prominent examples include PalmTree~\cite{li2021palmtree} and Trex~\cite{pei2020trex}, with the latter applying the Transformers on so-called micro-traces of the execution. However, these models are significantly outperformed by jTrans~\cite{wang2022jtrans} and CLAP~\cite{wang2024clap}, the Transformer-based backbone models investigated in this work.

Due to the structural nature of software code that can be represented as graphs, the use of GNNs has been explored in various forms as well. The control flow graph (CFG) between single instructions or linear basic blocks is leveraged in various architectures including Gemini~\cite{xu2017neural}, Order Matters~\cite{yu2020order}, and VulHawk~\cite{luo2023vulhawk}, as well as in experiments by \citeauthor{massarelli2019investigating}~\cite{massarelli2019investigating} on how to extract features from the CFG. Some models incorporate graph features in addition to the CFG. For instance, VulSeeker~\cite{gao2018vulseeker} integrates the data flow graph (DFG) and XBA~\cite{kim2022improving} also models information such as external function calls and references to string literals.

However, the key graph component investigated in this work is the call graph, in which nodes are represented by functions rather than lower-level sections of the code. While early work by \citeauthor{liu2018alphadiff} on $\alpha$Diff~\cite{liu2018alphadiff} leverages information from the call graph, it boils down to the functions' in- and out-degrees only. In more recent years, models such as BMM~\cite{guo2022exploring} and CFG2VEC~\cite{yu2023cfg2vec} were presented, which process call graphs alongside CFGs and---in the case of BMM---DFGs. Lastly, BinEnhance~\cite{wang2025binenhance} acts as a framework that improves embeddings generated by a pretrained backbone. It aggregates the call graph into a custom graph with additional edges representing data-co-use, address-adjacency, and string-use, and passes this graph through a GNN. It is important to note that our goal is not to outperform any of these approaches. For instance, the authors of BinEnhance argue that the call graph is insufficient to fully leverage inter-function semantics~\cite{wang2025binenhance}. Instead, this work specifically aims to carve out and comprehend the role of the call graph, hence understanding in which cases the usage of call graph-dependent GNNs is particularly beneficial.

The task of predicting function names is closely related to BCSD, though less commonly used as a downstream task to evaluate the generalizability of function embeddings. Of the models just discussed, only CFG2VEC~\cite{yu2023cfg2vec} is evaluated using a retrieval-based methodology for function name prediction. While earlier landmark papers in this domain, such as Debin~\cite{he2018debin}, use a probabilistic model on a dependency graph, more recent work relies on creating meaningful semantic embeddings for better performance. Their embedding models are conceptualized as encoders for this task in much the same way as for BCSD, and are combined with some sort of decoder model that performs the prediction which is commonly set up as a multilabel classification task. We use XFL's~\cite{patrick2023xfl} tree-based classifier in our work. The authors present DEXTER as its function embedding model which focuses on encoding features obtained via static analysis, including information from the call graph such as the number of the function's callers and callees. Other noteworthy approaches to function name prediction opt for deep learning-based decoders. For instance, SymLM~\cite{jin2022symlm} encodes functions using the pretrained Trex~\cite{pei2020trex} model. It then decodes the function name with an MLP by concatenating embeddings from the target function as well as its most frequent callers and callees, thus also leveraging information from the call graph. NERO~\cite{david2020neural} and BLens~\cite{benoit2025blens}, on the other hand, employ Transformer-based decoders. NERO augments the CFG with the calling context using static analysis for its encoder, whereas BLens attempts to consolidate previous advancements in the areas of BCSD and function name prediction by applying an ensemble encoder that leverages both PalmTree~\cite{li2021palmtree} and CLAP~\cite{wang2024clap} as well as DEXTER~\cite{patrick2023xfl}.

Compiler provenance recovery, in turn, focuses on technical details of the code rather than its semantic meaning, e.g., the classification of compiler optimization levels. While this task is used to evaluate BMM~\cite{guo2022exploring}, showing that task-specific training can indeed leverage the call graph as well, the most prominent work in this area is mostly decoupled from semantic tasks due to their contradictory goals as we found in \autoref{sec:results_cod}. For instance, an early landmark paper by \citeauthor{rosenblum2011recovering}~\cite{rosenblum2011recovering} uses support vector machines (SVMs) and conditional random fields (CRFs) on features extracted from the CFG to predict the compiler family and version as well as the optimization, surpassing an accuracy of 97\% on the latter classification given their dataset. More recent work leverages deep learning models including Transformer-based encoders in the case of BinProv~\cite{he2022binprov}, which employs a lightweight classification head to predict the compiler type and optimization level. GNNs have been employed in this area as well. Apart from \citeauthor{massarelli2019investigating}~\cite{massarelli2019investigating}, who also examine compiler family prediction in their experiments on extracting features from the CFG, Vestige~\cite{ji2021vestige} is a recent approach that applies a GAT on an attributed call graph to reconstruct the compilation provenance, demonstrating similarities to the models we analyze in this paper.

To our knowledge, only a limited amount of literature exists in the area of binary analysis that focuses on the understanding of binary function embedding models beyond conventional ablation studies. Most notably, previous work~\cite{dannehl2025instructions,xu2023improving} has leveraged explainability methods to investigate the importance of different instruction types on the function embedding. This is done by systematically applying occlusion-based saliency methods~\cite{bastings2020elephant}, an approach that---in the context of binary functions---assesses instruction importance by measuring how much the embedding changes after concealing it in the input. In our case, we focus on investigating the model performance in different cases rather than the importance of specific parts of the input. Therefore, we turn to an approach found in various application areas, for example to detect racial biases in algorithms~\cite{obermeyer2019dissecting}, in which the dataset is sliced along domain-specific dimensions in order to better understand the behavior of machine learning models.

\section{Conclusion}

In this work, we investigate the impact of enhancing binary function embeddings with context from the call graph. Training several graph-based models on binary code similarity detection, we find that even the best-performing models do not necessarily generalize to a function labeling task. Moreover, our results suggest that optimizing for such semantic understanding will typically deteriorate model performance on syntactic tasks such as compiler optimization detection. By examining the performance of models along domain-specific dimensions, we show that GNN-generated embeddings are able to effectively aggregate larger amounts of contextual information found in the call graph, hence making them more robust in a number of scenarios in which the initial embedding models demonstrated more significant performance drops. We hope that our research motivates model developers to further explore how inter-function information can best be leveraged in binary analysis and which pretraining tasks allow for the best generalizability toward rather semantic or syntactic tasks, particularly given the inherent trade-off between semantic similarity and syntactic nuance identified in this paper. Finally, we advocate for the wider adoption of similar explanatory analyses to identify and address both shortcomings and strengths of emerging models.

\begin{acks}
We would like to thank Moritz Dannehl for developing the initial preprocessing pipeline that formed the foundation of our experimental setup. Additionally, we extend our gratitude to the anonymous reviewers for their helpful comments that improved the clarity of this paper.
\end{acks}

\bibliographystyle{ACM-Reference-Format}
\bibliography{bibliography}

@inproceedings{guo2022exploring,
author = {Guo, Yixin and Li, Pengcheng and Luo, Yingwei and Wang, Xiaolin and Wang, Zhenlin},
title = {Exploring GNN based program embedding technologies for binary related tasks},
year = {2022},
isbn = {9781450392983},
publisher = {Association for Computing Machinery},
address = {New York, NY, USA},
url = {https://doi.org/10.1145/3524610.3527900},
doi = {10.1145/3524610.3527900},
booktitle = {Proceedings of the 30th IEEE/ACM International Conference on Program Comprehension},
pages = {366–377},
numpages = {12},
location = {Virtual Event},
series = {ICPC '22}
}

@inproceedings{wang2025binenhance,
  title={BinEnhance: An enhancement framework based on external environment semantics for binary code search},
  author={Wang, Yongpan and Li, Hong and Zhu, Xiaojie and Li, Siyuan and Dong, Chaopeng and Yang, Shouguo and Qin, Kangyuan},
  booktitle={32nd Annual Network and Distributed System Security Symposium},
  location = {San Diego, CA, USA},
  venue = {NDSS},
  series = {NDSS '25},
  year={2025}
}

@inproceedings{patrick2023xfl,
  author={Patrick-Evans, James and Dannehl, Moritz and Kinder, Johannes},
  booktitle={2023 IEEE Symposium on Security and Privacy (S\&P)}, 
  title={XFL: Naming functions in binaries with extreme multi-label learning}, 
  year={2023},
  pages={2375-2390},
  publisher={IEEE},
  doi={10.1109/SP46215.2023.10179439}
}

@inproceedings{wang2024clap,
author = {Wang, Hao and Gao, Zeyu and Zhang, Chao and Sha, Zihan and Sun, Mingyang and Zhou, Yuchen and Zhu, Wenyu and Sun, Wenju and Qiu, Han and Xiao, Xi},
title = {CLAP: Learning transferable binary code representations with natural language supervision},
year = {2024},
isbn = {9798400706127},
publisher = {Association for Computing Machinery},
address = {New York, NY, USA},
url = {https://doi.org/10.1145/3650212.3652145},
doi = {10.1145/3650212.3652145},
booktitle = {Proceedings of the 33rd ACM SIGSOFT International Symposium on Software Testing and Analysis},
pages = {503–515},
numpages = {13},
location = {Vienna, Austria},
series = {ISSTA '24}
}

@inproceedings{wang2022jtrans,
author = {Wang, Hao and Qu, Wenjie and Katz, Gilad and Zhu, Wenyu and Gao, Zeyu and Qiu, Han and Zhuge, Jianwei and Zhang, Chao},
title = {jTrans: Jump-aware transformer for binary code similarity detection},
year = {2022},
isbn = {9781450393799},
publisher = {Association for Computing Machinery},
address = {New York, NY, USA},
url = {https://doi.org/10.1145/3533767.3534367},
doi = {10.1145/3533767.3534367},
booktitle = {Proceedings of the 31st ACM SIGSOFT International Symposium on Software Testing and Analysis},
pages = {1–13},
numpages = {13},
location = {Virtual, South Korea},
series = {ISSTA '22}
}

@software{ida_pro,
  title        = {IDA Pro},
  organization = {Hex-Rays},
  version      = {7.6},
  year         = {2021},
  url          = {https://www.hex-rays.com/ida-pro}
}

@article{oord2018representation,
  title={Representation learning with contrastive predictive coding},
  author={Oord, Aaron van den and Li, Yazhe and Vinyals, Oriol},
  journal={arXiv preprint arXiv:1807.03748},
  year={2018}
}

@InProceedings{chen2020simple,
  title = {A simple framework for contrastive learning of visual representations},
  author = {Chen, Ting and Kornblith, Simon and Norouzi, Mohammad and Hinton, Geoffrey},
  booktitle = {Proceedings of the 37th International Conference on Machine Learning},
  pages = {1597--1607},
  year = {2020},
  volume = {119},
  series = {Proceedings of Machine Learning Research},
  month = {13--18 Jul},
  publisher = {PMLR},
  url = {https://proceedings.mlr.press/v119/chen20j.html},
}

@article{silva2020exploringsimclr,
  title={Exploring SimCLR: A simple framework for contrastive learning of visual representations},
  author={Silva, Thalles Santos},
  journal={https://sthalles.github.io},
  year={2020},
  url={https://sthalles.github.io/simple-self-supervised-learning/}
}

@inproceedings{hagberg2008networkx,
    title={Exploring network structure, dynamics, and function using NetworkX},
    author={Hagberg, Aric A and Schult, Daniel A and Swart, Pieter J},
    booktitle={Proceedings of the 7th Python in Science Conference},
    pages={11--15},
    year={2008},
    organization={SciPy},
    address={Pasadena, CA},
}

@article{comanici2025gemini,
  title={Gemini 2.5: Pushing the frontier with advanced reasoning, multimodality, long context, and next generation agentic capabilities},
  author = {Comanici, Gheorghe and Bieber, Eric and Schaekermann, Mike and Pasupat, Ice and Sachdeva, Noveen and Dhillon, Inderjit and Blistein, Marcel and Ram, Ori and Zhang, Dan and Rosen, Evan and Marris, Luke and Petulla, Sam and Gaffney, Colin and Aharoni, Asaf and Lintz, Nathan and Pais, Tiago and Jacobsson, Henrik and Szpektor, Idan and Jiang, Nan-Jiang and Hahn, Chris},
  journal={arXiv preprint arXiv:2507.06261},
  year={2025}
}

@misc{claude_sonnet_4_news,
  title = {Claude Sonnet 4},
  author = {{Anthropic}},
  howpublished = {\url{https://www.anthropic.com/news/claude-4}},
  year = {2025},
  note = {Accessed: 2025-08-18}
}

@misc{deep_seek_r1_0528,
  title = {{DeepSeek-R1-0528}},
  author = {{DeepSeek AI}},
  howpublished = {\url{https://huggingface.co/deepseek-ai/DeepSeek-R1-0528}},
  year = {2025},
  note = {Accessed: 2025-09-20}
}

@article{obermeyer2019dissecting,
author = {Ziad Obermeyer  and Brian Powers  and Christine Vogeli  and Sendhil Mullainathan },
title = {Dissecting racial bias in an algorithm used to manage the health of populations},
journal = {Science},
volume = {366},
number = {6464},
pages = {447-453},
year = {2019},
doi = {10.1126/science.aax2342},
URL = {https://www.science.org/doi/abs/10.1126/science.aax2342},
}

@inproceedings{dannehl2025instructions,
  author={Dannehl, Moritz and Valenzuela, Samuel and Kinder, Johannes},
  booktitle={2025 IEEE Security and Privacy Workshops (SPW)}, 
  title={Which instructions matter the most: A saliency analysis of binary function embedding models}, 
  year={2025},
  pages={145-151},
  doi={10.1109/SPW67851.2025.00019}
}

@inproceedings{benoit2025blens,
author = {Benoit, Tristan and Wang, Yunru and Dannehl, Moritz and Kinder, Johannes},
title = {BLens: Contrastive captioning of binary functions using ensemble embedding},
year = {2025},
isbn = {978-1-939133-52-6},
publisher = {USENIX Association},
address = {USA},
booktitle = {Proceedings of the 34th USENIX Conference on Security Symposium},
articleno = {353},
numpages = {20},
location = {Seattle, WA, USA},
series = {SEC '25}
}

@inproceedings{xu2023improving,
author = {Xu, Xiangzhe and Feng, Shiwei and Ye, Yapeng and Shen, Guangyu and Su, Zian and Cheng, Siyuan and Tao, Guanhong and Shi, Qingkai and Zhang, Zhuo and Zhang, Xiangyu},
title = {Improving binary code similarity transformer models by semantics-driven instruction deemphasis},
year = {2023},
isbn = {9798400702211},
publisher = {Association for Computing Machinery},
address = {New York, NY, USA},
url = {https://doi.org/10.1145/3597926.3598121},
doi = {10.1145/3597926.3598121},
booktitle = {Proceedings of the 32nd ACM SIGSOFT International Symposium on Software Testing and Analysis},
pages = {1106–1118},
numpages = {13},
location = {Seattle, WA, USA},
series = {ISSTA '23}
}

@inproceedings{bastings2020elephant,
    title = "The elephant in the interpretability room: Why use attention as explanation when we have saliency methods?",
    author = "Bastings, Jasmijn  and Filippova, Katja",
    booktitle = "Proceedings of the Third BlackboxNLP Workshop on Analyzing and Interpreting Neural Networks for NLP",
    month = nov,
    year = "2020",
    address = "Online",
    publisher = "Association for Computational Linguistics",
    url = "https://aclanthology.org/2020.blackboxnlp-1.14/",
    doi = "10.18653/v1/2020.blackboxnlp-1.14",
    pages = "149--155",
}

@INPROCEEDINGS{ding2019asm2vec,
  author={Ding, Steven H. H. and Fung, Benjamin C. M. and Charland, Philippe},
  booktitle={2019 IEEE Symposium on Security and Privacy (SP)}, 
  title={Asm2Vec: Boosting static representation robustness for binary clone search against code obfuscation and compiler optimization}, 
  year={2019},
  pages={472-489},
  doi={10.1109/SP.2019.00003}
}

@InProceedings{massarelli2019safe,
author="Massarelli, Luca
and Di Luna, Giuseppe Antonio
and Petroni, Fabio
and Baldoni, Roberto
and Querzoni, Leonardo",
title="SAFE: Self-attentive function embeddings for binary similarity",
booktitle="Detection of Intrusions and Malware, and Vulnerability Assessment",
year="2019",
publisher="Springer International Publishing",
address="Cham",
pages="309--329",
isbn="978-3-030-22038-9"
}

@article{mikolov2013efficient,
  title={Efficient estimation of word representations in vector space},
  author={Mikolov, Tomas and Chen, Kai and Corrado, Greg and Dean, Jeffrey},
  journal={arXiv preprint arXiv:1301.3781},
  year={2013}
}

@inproceedings{vaswani2017attention,
author = {Vaswani, Ashish and Shazeer, Noam and Parmar, Niki and Uszkoreit, Jakob and Jones, Llion and Gomez, Aidan N. and Kaiser, \L{}ukasz and Polosukhin, Illia},
title = {Attention is all you need},
year = {2017},
isbn = {9781510860964},
publisher = {Curran Associates Inc.},
address = {Red Hook, NY, USA},
booktitle = {Proceedings of the 31st International Conference on Neural Information Processing Systems},
pages = {6000–6010},
numpages = {11},
location = {Long Beach, CA, USA},
series = {NIPS'17}
}

@inproceedings{li2021palmtree,
author = {Li, Xuezixiang and Qu, Yu and Yin, Heng},
title = {PalmTree: Learning an assembly language model for instruction embedding},
year = {2021},
isbn = {9781450384544},
publisher = {Association for Computing Machinery},
address = {New York, NY, USA},
url = {https://doi.org/10.1145/3460120.3484587},
doi = {10.1145/3460120.3484587},
booktitle = {Proceedings of the 2021 ACM SIGSAC Conference on Computer and Communications Security},
pages = {3236–3251},
numpages = {16},
location = {Virtual Event, Republic of Korea},
series = {CCS '21}
}

@ARTICLE{pei2020trex,
  author={Pei, Kexin and Xuan, Zhou and Yang, Junfeng and Jana, Suman and Ray, Baishakhi},
  journal={IEEE Transactions on Software Engineering}, 
  title={Learning approximate execution semantics from traces for binary function similarity}, 
  year={2023},
  volume={49},
  number={4},
  pages={2776-2790},
  doi={10.1109/TSE.2022.3231621}
}

@inproceedings{xu2017neural,
author = {Xu, Xiaojun and Liu, Chang and Feng, Qian and Yin, Heng and Song, Le and Song, Dawn},
title = {Neural network-based graph embedding for cross-platform binary code similarity detection},
year = {2017},
isbn = {9781450349468},
publisher = {Association for Computing Machinery},
address = {New York, NY, USA},
url = {https://doi.org/10.1145/3133956.3134018},
doi = {10.1145/3133956.3134018},
booktitle = {Proceedings of the 2017 ACM SIGSAC Conference on Computer and Communications Security},
pages = {363–376},
numpages = {14},
location = {Dallas, TX, USA},
series = {CCS '17}
}

@article{yu2020order,
title={Order Matters: Semantic-aware neural networks for binary code similarity detection},
volume={34},
url={https://ojs.aaai.org/index.php/AAAI/article/view/5466},
DOI={10.1609/aaai.v34i01.5466},
number={01},
journal={Proceedings of the AAAI Conference on Artificial Intelligence},
author={Yu, Zeping and Cao, Rui and Tang, Qiyi and Nie, Sen and Huang, Junzhou and Wu, Shi},
year={2020},
month={Apr.},
pages={1145-1152}
}

@inproceedings{luo2023vulhawk,
  title={VulHawk: Cross-architecture vulnerability detection with entropy-based binary code search},
  author={Luo, Zhenhao and Wang, Pengfei and Wang, Baosheng and Tang, Yong and Xie, Wei and Zhou, Xu and Liu, Danjun and Lu, Kai},
  booktitle={30th Annual Network and Distributed System Security Symposium},
  location = {San Diego, CA, USA},
  venue = {NDSS},
  series = {NDSS '23},
  year={2023},
  doi={10.14722/ndss.2023.24415}
}

@INPROCEEDINGS{gao2018vulseeker,
  author={Gao, Jian and Yang, Xin and Fu, Ying and Jiang, Yu and Sun, Jiaguang},
  booktitle={2018 33rd IEEE/ACM International Conference on Automated Software Engineering (ASE)}, 
  title={VulSeeker: A semantic learning based vulnerability seeker for cross-platform binary}, 
  year={2018},
  pages={896-899},
  doi={10.1145/3238147.3240480}
}

@inproceedings{kim2022improving,
author = {Kim, Geunwoo and Hong, Sanghyun and Franz, Michael and Song, Dokyung},
title = {Improving cross-platform binary analysis using representation learning via graph alignment},
year = {2022},
isbn = {9781450393799},
publisher = {Association for Computing Machinery},
address = {New York, NY, USA},
url = {https://doi.org/10.1145/3533767.3534383},
doi = {10.1145/3533767.3534383},
booktitle = {Proceedings of the 31st ACM SIGSOFT International Symposium on Software Testing and Analysis},
pages = {151–163},
numpages = {13},
location = {Virtual, South Korea},
series = {ISSTA '22}
}

@inproceedings{liu2018alphadiff,
author = {Liu, Bingchang and Huo, Wei and Zhang, Chao and Li, Wenchao and Li, Feng and Piao, Aihua and Zou, Wei},
title = {$\alpha$Diff: Cross-version binary code similarity detection with DNN},
year = {2018},
isbn = {9781450359375},
publisher = {Association for Computing Machinery},
address = {New York, NY, USA},
url = {https://doi.org/10.1145/3238147.3238199},
doi = {10.1145/3238147.3238199},
booktitle = {Proceedings of the 33rd ACM/IEEE International Conference on Automated Software Engineering},
pages = {667–678},
numpages = {12},
location = {Montpellier, France},
series = {ASE '18}
}

@inproceedings{yu2023cfg2vec,
author = {Yu, Shih-Yuan and Achamyeleh, Yonatan Gizachew and Wang, Chonghan and Kocheturov, Anton and Eisen, Patrick and Faruque, Mohammad Abdullah Al},
title = {CFG2VEC: Hierarchical graph neural network for cross-architectural software reverse engineering},
year = {2023},
isbn = {9798350300376},
publisher = {IEEE Press},
url = {https://doi.org/10.1109/ICSE-SEIP58684.2023.00031},
doi = {10.1109/ICSE-SEIP58684.2023.00031},
booktitle = {Proceedings of the 45th International Conference on Software Engineering: Software Engineering in Practice},
pages = {281–291},
numpages = {11},
location = {Melbourne, Australia},
series = {ICSE-SEIP '23}
}

@inproceedings{he2018debin,
author = {He, Jingxuan and Ivanov, Pesho and Tsankov, Petar and Raychev, Veselin and Vechev, Martin},
title = {Debin: Predicting debug information in stripped binaries},
year = {2018},
isbn = {9781450356930},
publisher = {Association for Computing Machinery},
address = {New York, NY, USA},
url = {https://doi.org/10.1145/3243734.3243866},
doi = {10.1145/3243734.3243866},
booktitle = {Proceedings of the 2018 ACM SIGSAC Conference on Computer and Communications Security},
pages = {1667–1680},
numpages = {14},
location = {Toronto, Canada},
series = {CCS '18}
}

@inproceedings{jin2022symlm,
author = {Jin, Xin and Pei, Kexin and Won, Jun Yeon and Lin, Zhiqiang},
title = {SymLM: Predicting function names in stripped binaries via context-sensitive execution-aware code embeddings},
year = {2022},
isbn = {9781450394505},
publisher = {Association for Computing Machinery},
address = {New York, NY, USA},
url = {https://doi.org/10.1145/3548606.3560612},
doi = {10.1145/3548606.3560612},
booktitle = {Proceedings of the 2022 ACM SIGSAC Conference on Computer and Communications Security},
pages = {1631–1645},
numpages = {15},
location = {Los Angeles, CA, USA},
series = {CCS '22}
}

@article{david2020neural,
author = {David, Yaniv and Alon, Uri and Yahav, Eran},
title = {Neural reverse engineering of stripped binaries using augmented control flow graphs},
year = {2020},
issue_date = {November 2020},
publisher = {Association for Computing Machinery},
address = {New York, NY, USA},
volume = {4},
number = {OOPSLA},
url = {https://doi.org/10.1145/3428293},
doi = {10.1145/3428293},
journal = {Proceedings of the ACM on Programming Languages},
month = nov,
articleno = {225},
numpages = {28}
}

@inproceedings{rosenblum2011recovering,
author = {Rosenblum, Nathan and Miller, Barton P. and Zhu, Xiaojin},
title = {Recovering the toolchain provenance of binary code},
year = {2011},
isbn = {9781450305624},
publisher = {Association for Computing Machinery},
address = {New York, NY, USA},
url = {https://doi.org/10.1145/2001420.2001433},
doi = {10.1145/2001420.2001433},
booktitle = {Proceedings of the 2011 International Symposium on Software Testing and Analysis},
pages = {100–110},
numpages = {11},
location = {Toronto, Ontario, Canada},
series = {ISSTA '11}
}

@inproceedings{he2022binprov,
author = {He, Xu and Wang, Shu and Xing, Yunlong and Feng, Pengbin and Wang, Haining and Li, Qi and Chen, Songqing and Sun, Kun},
title = {BinProv: Binary code provenance identification without disassembly},
year = {2022},
isbn = {9781450397049},
publisher = {Association for Computing Machinery},
address = {New York, NY, USA},
url = {https://doi.org/10.1145/3545948.3545956},
doi = {10.1145/3545948.3545956},
booktitle = {Proceedings of the 25th International Symposium on Research in Attacks, Intrusions and Defenses},
pages = {350–363},
numpages = {14},
location = {Limassol, Cyprus},
series = {RAID '22}
}

@inproceedings{massarelli2019investigating,
  title={Investigating graph embedding neural networks with unsupervised features extraction for binary analysis},
  author={Massarelli, Luca and Di Luna, Giuseppe Antonio and Petroni, Fabio and Querzoni, Leonardo and Baldoni, Roberto and others},
  booktitle={Proceedings BAR 2019 Workshop on Binary Analysis Research},
  pages={1--11},
  year={2019},
  doi={10.14722/bar.2019.23020}
}

@inproceedings{ji2021vestige,
author = {Ji, Yuede and Cui, Lei and Huang, H. Howie},
title = {Vestige: Identifying binary code provenance for vulnerability detection},
year = {2021},
isbn = {978-3-030-78374-7},
publisher = {Springer-Verlag},
address = {Berlin, Heidelberg},
url = {https://doi.org/10.1007/978-3-030-78375-4_12},
doi = {10.1007/978-3-030-78375-4_12},
booktitle = {Applied Cryptography and Network Security: 19th International Conference, ACNS 2021, Kamakura, Japan, June 21–24, 2021, Proceedings, Part II},
pages = {287–310},
numpages = {24},
location = {Kamakura, Japan}
}

@inproceedings{velivckovic2017graph,
  author={Veli{\v{c}}kovi{\'c}, Petar and Cucurull, Guillem and Casanova, Arantxa and Romero, Adriana and Li{\'o}, Pietro and Bengio, Yoshua},
  title        = {Graph attention networks},
  booktitle    = {Proceedings of the 6th International Conference on Learning Representations},
  location = {Vancouver, BC, Canada},
  series = {ICLR '18},
  venue = {ICLR},
  year         = 2018,
}

@inproceedings{kipf2017semi,
  author = {Kipf, Thomas N. and Welling, Max},
  booktitle = {Proceedings of the 5th International Conference on Learning Representations},
  location = {Palais des Congr{\`e}s Neptune, Toulon, France},
  series = {ICLR '17},
  title = {{Semi-supervised classification with graph convolutional networks}},
  venue = {ICLR},
  year = 2017
}

@inproceedings{raff2018malware,
  title={Malware detection by eating a whole EXE},
  author={Edward Raff and Jon Barker and Jared Sylvester and Robert Brandon and Bryan Catanzaro and Charles K. Nicholas},
  booktitle={The Workshops of the Thirty-Second AAAI Conference on Artificial Intelligence},
  location = {New Orleans, LA, USA},
  series       = {{AAAI} Technical Report},
  volume       = {{WS-18}},
  pages        = {268--276},
  publisher    = {{AAAI} Press},
  year         = {2018}
}

\end{document}